\documentclass[12pt]{article}
\usepackage{graphicx}
\begin{document}
\title{A REMARKABLE ANGULAR DISTRIBUTION OF THE INTERMEDIATE
SUBCLASS OF THE GAMMA-RAY BURSTS}
\author{
ATTILA M\'ESZ\'AROS$^{1, 2, 3}$ \and ZSOLT 
BAGOLY$^4$
\and ISTV\'AN HORV\'ATH$^5$ 
\and LAJOS G. BAL\'AZS$^2$ 
\and and ROLAND VAVREK$^{2,6}$
}
\maketitle

\bigskip

$^1$Department of Astronomy, Charles University, V
Hole\v{s}ovi\v{c}k\'ach 2, CZ-180 00 Prague 8, Czech Republic

$^2$Konkoly Observatory, Box 67, H-1505 Budapest, Hungary

$^3$Department of Astronomy, E\"{o}tv\"{o}s University, P\'azm\'any P\'eter
s\'et\'any 1/A, H-1518 Budapest, Hungary

$^4$Laboratory for Information Technology,
E\"{o}tv\"{o}s University, P\'azm\'any P\'eter
s\'et\'any 1/A, H-1518 Budapest, Hungary

$^5$Department of Physics, BJKMF, Box 12, H-1456 Budapest, Hungary

$^6$Observatoire de Paris, F-92195 Meudon Cedex, Prance

\section*{ABSTRACT}

In the article a test is developed, which allows to test the null-hypothesis
of the intrinsic randomness in the angular distribution
of gamma-ray bursts collected at the Current BATSE Catalog. The method
is a modified version of the well-known counts-in-cells test, and
fully eliminates the non-uniform sky-exposure function of BATSE instrument.
Applying this method to the case
of all gamma-ray bursts no intrinsic non-randomness was found. The test also
did not find intrinsic non-randomnesses for the short
and long gamma-ray bursts, respectively.
On the other hand, using the method to the new intermediate subclass of
gamma-ray bursts, the null-hypothesis of the
intrinsic randomness for 181 intermediate gamma-ray bursts is
rejected on the $96.4\%$ confidence level. Taking 92 dimmer bursts from
this subclass itself, we obtain the 
surprising result: This "dim" subclass 
of the intermediate subclass has an intrinsic non-randomness on the
99.3\% confidence level. On the other hand, the 89 "bright" GRBs show no 
intrinsic non-randomness.\\
{\it Subject headings:} cosmology: observations - gamma rays: bursts

\section{INTRODUCTION}

Two results of the last years in the statistics of the gamma-ray
bursts (GRBs) are doubtlessly remarkable.

The first one concerns the number of subclasses.
Recently, two different articles (\cite{muk}; \cite{hor})
simultaneously suggest that the earlier separation \cite{kou}
of GRBs into short and long subclasses is incomplete.
(It is a common practice to call GRBs having $T_{90} <2$ s ($T_{90}
> 2$ s) as short (long) GRBs, where $T_{90}$ is the time during which
90\% of the fluence is accumulated
\cite{kou}.) These articles show that,
in essence, the earlier long subclass alone should be further separated
into a new "intermediate" subclass ($2$ s $< T_{90} < 10$ s) and into
a "truncated long" subclass ($T_{90} > 10$ s). 
(In what follows, the long subclass will contain only the
GRBs with $T_{90} > 10$ s, and the intermediate subclass will be
considered as a new subclass.)

The second result concerns the angular distribution of GRBs.
At the last years several attempts (\cite{hart}, \cite{briggs}; $\;$
\cite{teg}; [Bal\'azs et al. 1998]; 
\cite{ba99}) were done either
to confirm or to reject the randomness in the angular sky distribution
of GRBs being collected at BATSE Catalog (\cite{fish}; \cite{mee}).
Theoretically, if the intrinsic distribution
of GRBs is actually random, an observation of some non-randomness
is still expected
due to the BATSE non-uniform sky-exposure function (\cite{fish};
\cite{mee}). Hartmann et al. (1991),
 Briggs et al. (1996) and Tegmark et al. (1996b)
did not find any statistically significant departure from
the randomness. On the other hand, the existence of {\it some} non-randomness
was confirmed on the $>99.9\%$ confidence level by Bal\'azs et al. (1998). 
This behavior can be caused either purely by instrumental effects or
the instrumental effects alone do not explain fully the detected behavior
and some intrinsic non-randomnesses should also exist.  
Bal\'azs et al. (1998, 1999) suggest the second possibility.  
This conclusion follows from the result that while the short subclass
shows a non-randomness, the intermediate + long subclasses do not indicate it.
It is difficult to explain such behavior of subclasses by the instrumental
effects alone.

In this article we will again investigate the angular distribution of
GRBs. Trivially, after the discovery of the new intermediate subclass,
it is highly required to test the intrinsic randomness in the angular
distribution of this new subclass, too. In addition, of course,
new different tests, which
exactly eliminate the effect of the sky-exposure function, are also
required in order to complete the results of 
Bal\'azs et al. (1998, 1999).

The aim of this article is to
test the {\it intrinsic} randomness in the angular distribution 
of all GRBs and of the three subclasses separately, too. 
We will use a modification of the well-known
counts-in-cells method. This is a standard and simple statistical test
(see, cf., M\'esz\'aros (1997) 
and references therein). The advantage of this method is
given by the fact that 
it allows to eliminate quite simply and exactly the sky-exposure function. 
The main result of paper will be the surprising
conclusion that the intermediate subclass and only this subclass alone
suggests a non-randomness on the 96.4\% confidence level; its "dimmer"
half even on the 99.3\% confidence level.

The paper is organized as follows. In Section 2 the method is 
described. In Section 3 the results of test 
are presented. Section 4 discusses and summarizes
the results of the article.

\section{THE TEST}

Assume for the moment that there is no non-uniform sky-exposure
function. We separate the sky in declination into $m_{dec} > 1$ stripes
having the same area ($4\pi/m_{dec}$ steradian). The boundaries
of stripes are the declinations $\delta_k, \;k=0,1,...,m_{dec}$, where
$\delta_0 = -90$ degree and $\delta_{m_{dec}} = +90$ degree, respectively.
The remaining values are analytically calculable, and appear symmetrically
with respect to $\delta =0$. One has:
$\sin \delta_k = 2k/m_{dec} -1$. (For example: If $m_{dec} =3$, then
$\delta_{1,2}= \pm 19.47$ degree; if $m_{dec} =4$, then
$\delta_{1,3}= \pm 30.00$ degree and $\delta_2= 0.00$ degree; etc.)
We also separate the sky in right ascension $\alpha$ into
$m_{ra}>1$ stripes. They are
defined by boundaries $\alpha = 360 k'/m_{ra}$ degree; $k'= 0,1,..,m_{ra}$.
(Obviously, a trivial modification of this separation in right ascension
is the case, when the boundaries are $\alpha = 360 (k'+ p)/m_{ra}$ degree,
where $p$ is an arbitrary real number fulfilling $0 <p < 1$.)
All this means that we separated the sky into $M = m_{dec}\times m_{ra}$
areas ("cells") having the same size $4\pi/M$ steradian. If there are
$N$ GRBs on the sky, then $n =N/M$ is the mean of GRBs 
at a cell. Let $n_i;\; i=1,2,...,M$ be the observed number of GRBs at the
$i$th cell ($\sum_{i=1}^{M} n_i = N$). Then
\begin{equation}
var_M = (M-1)^{-1}\sum_{i=1}^{M} (n_i -n)^2
\end{equation}
defines the observed variance. For the 
given cell structure with $M$ cells,
 due to the Bernoulli distribution
(\cite{me97}, \cite{ba98}), the measured
variance $var_M$
should be identical to the theoretically expected value
$n(1-1/M)$. This theoretical prediction should then be tested.

Note that this and similar methods (see, e.g., M\'esz\'aros (1997)
for details and further references) are usual in astronomy.
For example,
this method was used already by Abell (1958)
to reject the randomness
in the sky-distribution of clusters of galaxies. The test is, compared
with other statistical tests ("two-point angular correlation function",
"nearest neighbor distances", etc.; \cite{peebles}; \cite{diggle};
\cite{pasztor}), not the most sensitive one to detect non-randomnesses.
Its importance for our purposes is given by the fact that it allows an
extremely simple generalization to the case with non-zero sky-exposure
function. 

Now, we generalize the method to this case. This may easily be done
by changing the boundaries
of cells in order to have the same probability (and hence the same expected
number $n = N/M$) for a given cell. The sky-exposure function
is a function of declination only (\cite{fish}; \cite{mee}).
Hence, the choice
of equatorial coordinates is highly convenient, because then no changes of
boundaries are necessary in right ascension. The new boundaries $\delta_k,
\; k=0,1,...,m_{dec}$
in declination may be calculated analytically as follows. Clearly,
$\delta_{0,m_{dec}} = \pm 90$ degree remain. In BATSE Catalog
\cite{mee} the exposure function $f(\delta)$
is defined for 37 values of declination
(for $\delta_r =-90, -85, ..., +85, +90$ degree; $r=0,1,2,..,36$).
To obtain $\delta_k$ we, first, calculate
the value
\begin{equation}
A = \frac{5\pi}{180}\sum_{r=1}^{35} f(\delta_r) \cos \delta_r \;,
\end{equation}
where for the given $r$ the corresponding declination is
$\delta_r = -90 + 5r$ degree. (Remark that $r=0$ and $r=36$, respectively,
need not be in the sum, because $\cos (\pm 90) =0$.) Then, second,
for $m_{dec} \geq 2$ we search for the values $\delta_k;\;
k=1,2,...,(m_{dec}-1)$ as follows.
For the given $k$ we search for the declination
$\delta_i$ fulfilling the condition
\begin{equation}
\frac{\pi}{36A}\sum_{j=1}^{i} f(\delta_j) \cos \delta_j \leq
\frac{k}{m_{dec}} <
\frac{\pi}{36A}\sum_{j=1}^{i+1} f(\delta_j) \cos \delta_j\;.
\end{equation}
Having this we search, by linear interpolation between $\delta_i$ and
$\delta_{i+1}$, for the exact value of $\delta_k$.
By this method $\delta_k$ is well calculable. (For example: For
$m_{dec} = 3$ we obtain $\delta_1 = -19.51$ degree, $\delta_2 = 22.44$ degree;
for $m_{dec} = 4$ we obtain $\delta_1 = -30.83$ degree,
$\delta_2 = 1.51$ degree, $\delta_3 = 33.60$ degree; etc.)
Having these cells with these "shifted" boundaries in declination the
variance may be calculated identically to the case with no sky-exposure
function. This method will test the pure {\it intrinsic} randomness; 
the effect of BATSE sky-exposure function is exactly eliminated.

It is natural to probe
different values of $M$. In addition, for some $M$ different
cell structures are still possible (cf. $M =12$ allows
$m_{dec} =2,3,4,6$). Hence, generally,
several - say $Q$ - cell structures may be probed for the same sample of
GRBs. Having these $Q$ cell structures (and hence $Q$ means + $Q$ measured
variances) two questions arise. 1. How to
calculate the confidence level for a given cell structure?
2. Having $Q$ values of confidence levels, how to
calculate the final confidence level? The answer for the first question
seems to be quite clear: $var_M/n$ seems to be identical to the
$\chi^2$ value for $M-1$ degree of freedom (\cite{trumpler}; \cite{kendall};
\cite{press}; the mean is obtained from the sample itself, and therefore
the degree of freedom is $M-1$). Nevertheless, the situation is not so obvious,
because the $\chi^2$ test needs $n >5$ (\cite{trumpler}; \cite{kendall};
\cite{press}). In addition, some statistical text-books propose to use
"quadratic" cells only (\cite{diggle}, Chapt. 2.5.). If all these restrictions
were taken into account, then $\chi^2$ tests would be possible only for
$2 m_{dec} = m_{ra}; \; M = 2m_{dec}^2$ and $N > 5M = 10 m_{dec}^2$.
This would be a drastic truncation of the possible cell structures.
But, not doing these restrictions,
the estimation of the confidence level for a given cell structure
must be done by more complicated procedures; e.g., by numerical simulations.
Concerning the answer to the second question the situation is even
less clear.
As the reasonable search for the final confidence level only Monte Carlo
simulations seem to be usable \cite{press}.

Keeping all this in mind, we will proceed as follows. 
In the coordinate system with axes
$x = 1/M$ versus $y = \sqrt{var_M/n} = (var/mean)^{1/2}$
the $Q$ values of $(var/mean)^{1/2}$ define $Q$ points 
(one point for any cell structure; $y_j = \sqrt{var_{M,j}/n}$,
where $j=1,2,....,Q$). Clearly,
for these points one expects the theoretical curve $y = \sqrt{1 -x}$.
This theoretical expectation can straightforwardly be
verified, e.g., by least squares estimation (\cite{press}, Chapt. 15.2.;
\cite{diggle}, Chapt. 5.3.1).
Our estimator is the dispersion
\begin{equation}
\sigma_{Q} = \sum_{j=1}^{Q} (y_j - \sqrt{1-1/M}\;)^2\;.
\end{equation}
Obviously, smaller $\sigma_Q$ suggests that the theoretical curve is better
fitted. Note still that, as the best choice,
the square root of $var_M/n$ is proposed in this
"var/mean" test (\cite{diggle}, Chapt. 5.4.).

The confidence level can then be estimated by Monte Carlo simulations in the
following way: We throw 1000-times randomly $N$ 
points on the sphere, and repeat the above calculation leading to 
$\sigma_{Q}$ for every simulated sample. Then we compare the size of
the $\sigma_Q$ obtained from this simulation with $\sigma_Q$ obtained
from the actual GRB positions. Let $\omega$ be the number
of simulations, when the obtained $\sigma_{Q}$ is bigger than the
actual value of $\sigma_{Q}$.
Then one may conclude that $(100-\omega/10)$ is the confidence level
in percentage.
Clearly, this method does not need $n >5$ and quadratic cells. 
 
There
is no commonly accepted confidence level in statistics, 
above which the null-hypothesis
should already be rejected (\cite{trumpler}; \cite{kendall}). 
It is only a general agreement that confidence levels smaller than $95\%$
should not be considered. Our opinion is (see also \cite{kendall}) that
the confidence levels bigger than $95\%$ can already be taken as "remarkable",
"suspicious", "interesting", etc.; a higher than 99\% confidence
level may still mean the rejection of null-hypothesis, and such result must
doubtlessly be announced.
Hence, we will require that the confidence level be bigger than $95\%$.
Thus, here it must be $\omega < 50$. 

In this paper GRBs will be taken between trigger-numbers
 0105 and 6963 from Current BATSE Catalog 
\cite{mee} having defined $T_{90}$ (i.e. all GRBs detected up to August 1996
having measured $T_{90}$). From
them we exclude, similarly to Pendleton et al. (1997) and 
Bal\'azs et al. (1998), the faintest
GRBs having a peak flux (on 256 ms trigger) smaller than 0.65
photon/(cm$^2$s). This truncation is proposed by Pendleton et al. (1997)
 in order to
avoid the problems with the changing threshold.
The 1284 GRBs obtained in this way define the "all" class. From them
there were 339 GRBs with $T_{90} < 2$ s (the "short" subclass), 181 GRBs
with $2$ s $< T_{90} < 10$ s (the "intermediate" subclass) and 764 GRBs with
$T_{90} > 10$ s (the "long" subclass). We will study the all class and
the three subclasses separately. 

We will ad hoc choose $m_{dec} = 2, 3, \ldots, 8$
and $m_{ra} = 2, 3, \ldots, 16$. I.e. it will be $Q = 105$.
Of course, this choice of $Q$ is more or less subjective. Nevertheless,
our choice is motivated by two concrete 
arguments. First, we would like to study only
the angular scales much bigger than the positional errors.
(The size of a cell will not be smaller than 22.5 degree.
On these angular scales
no problems should arise from the positional errors \cite{mee}.)
Second, it is reasonable not to consider such high values of $m_{dec}$, when
$180/m_{dec}$ is already 
comparable or even smaller than 5 degree. (If this were
not required then the elimination of sky-exposure function would be
problematic due to its definition for declination intervals with widths
5 degree.)

\section{THE RESULTS}

Figure 1 collects the results of $Q=105$ "var/mean" tests of four
different cases. It is obvious immediately that for the "all"
case the points follow well the theoretical curve. For the "short"
and "long" subclasses, on the other hand, there is a slight tendency
of points to be above the theoretical curve. 
The situation concerning the intermediate
subclass seems to be the most unambiguous: mainly for small $M$ (roughly below
$M \simeq 40$) the points are clearly above the theoretical curve.
This suggests an intrinsic non-randomness mainly in the sky
distribution of intermediate subclass;
such possibility for the short and long subclasses, respectively,
cannot be excluded, too. 

The results of Monte Carlo simulations support this expectation only
in the case of intermediate subclass.
We obtain $\omega = 287$ ($\omega = 80$, $\omega = 36$, $\omega = 440$)
for all GRBs (short, intermediate, long GRBs). Hence,
the rejection of null-hypothesis is confirmed 
for the intermediate subclass only
on the 96.4\% confidence level. For the short and long subclasses, 
respectively, and also for
all GRBs the null-hypothesis cannot be rejected on the
$>95\%$ confidence level. For the short subclass we have a $92\%$
confidence level; for the remaining two cases even smaller levels.

\section{DISCUSSION AND CONCLUSION}

The most surprising result of paper concerns the intermediate subclass.
The intrinsic non-randomness is confirmed on
the confidence level $>95\%$.
This confidence level, as discussed in Section 2, is "remarkable", but
is not enough to reject the null-hypothesis of randomness.

The results concerning the $339$ short
GRBs should also be mentioned. Nevertheless, 
the $92\%$ confidence level 
is clearly not enough to reject the confidence null-hypothesis.
On the other hand, this result, 
together with Bal\'azs et al. (1998, 1999), suggest that also
for the short subclass itself the rejection of null-hypothesis
of intrinsic randomness can also occur by further tests.   

In the case of $764$ long GRBs, and also of the $1284$
all GRBs, there are no indications for the
non-randomnesses. All this seems to be in accordance with the results of
Bal\'azs et al. (1998, 1999).

We think that the result concerning the intermediate subclass 
is highly surprising, because
just this new subclass, having the smallest number of GRBs,
has a remarkable "proper" behavior. 

A short further investigation of this subclass fully supports this
conclusion.

There are 181 GRBs in this intermediate subclass. Be divided this
subclass into two further subclasses; into the "dim" and "bright" ones.
By chance the peak flux = 2 photons/(cm$^2$s) (on $0.256$s trigger) is 
practically
identical to the medium of peak flux for this subclass. Therefore, we consider
the GRBs having smaller (bigger) peak flux 2 photons/(cm$^2$s)
as the "dim" ("bright") subclass of the intermediate subclass. There
are 92 GRBs at the "dim" subclass, and 89 GRBs at the "bright" one.

We provide the 105 "var/mean" tests for these two parts, too. We obtain
the surprising result that the "dim" subclass has an intrinsic non-randomness
on the 99.3\% confidence level ($\omega = 7$). Contrary this, 
the "bright" subclass can still be random ($\omega = 662$).   

The sky distribution of $92$ intermediate dim GRBs is shown on Figure 2.

We mean that the behavior of the intermerdiate subclass of GRBs, 
quite independently,
supports the correctness of the introduction of this new subclass 
(\cite{muk}; \cite{hor}). Further 
investigations of this new subclass are highly required.

Three notes are still needed.

First, purely from the statistical point of view,
it must be precised
that even the rejection of null-hypothesis
of the intrinsic randomness would not mean a pure intrinsic non-randomness
in the spatial angular distribution of GRBs. This is given
by the fact that, up to now, it cannot be fully excluded that GRBs (or some
part of them) are not unique phenomenons, and there can occur some 
repetitions, too. This question is studied intensively by several 
papers (\cite{mee95}, \cite{qua95}, \cite{qua96}, \cite{tega}, \cite{graz},
\cite{hak}) con\-clu\-ding
that repetition can still play a role. 

Second, strictly speaking, the statistical 
counts-in-cells test is testing the "complete spatial randomness" (shortly
the "randomness") of the distribution of GRB on the celestial sphere 
(\cite{diggle}, Chapt.1.3). Therefore, in this
paper we have kept this terminology. In cosmology, on the other hand,
the word "random" ("non-random") is rarely used, and the word "isotropic" 
("anisotropic") is usual 
(for the exact definition of isotropy in cosmology see, e.g., 
Weinberg (1972), Chapt. 14.1). 
Of course, here we will not go into the details of these terminology questions
(see, e.g., Peebles (1980) for more details concerning these questions). 
We note only that the "random-isotropic" ("non-random-anisotropic") 
substitution is quite acceptable  
on the biggest angular scales; on smaller angular scales the
situation is not so clear. 
Therefore, in Bal\'azs et al. (1998, 1999), where only the
angular scales $\sim 90$ degrees and higher were studied, the words
"isotropy" and "anisotropy" were quite usable. 
In this article, going down up to the scales
$\sim (20-25)$ degree, the used terminology is more relevant.

Third, trivially, further studies are needed. They should
test - by other different statistical methods - again the intrinsic 
randomnesses (more generally: the intrinsic 
spatial distributions \cite{lamb}), 
both for all GRBs and for the subclasses. In addition,
a test of the repetition alone, i.e. a test not being influenced by positions,
is highly required.

As the conclusion, the results of paper may be summarized as follows.

\begin{itemize}

\item We developed a method, which can verify quite simply
the intrinsic randomness alone in the angular distribution of GRBs, because
the method eliminates exactly the non-zero sky-exposure function. 

\item We rejected the null-hypothesis of the intrinsic randomness
in the angular 
distribution of 181 intermediate GRBs on the $96.4\%$ confidence level.

\item We rejected the null-hypothesis of the intrinsic randomness
in the angular 
distribution of 92 "dim" intermediate GRBs on the $99.3\%$ confidence level.

\item We did not reject the null-hypotheses of the intrinsic randomnesses
in the angular distribution of the remaining two subclasses and of
the all GRBs, respectively, on the $>95\%$ confidence levels; the "bright"
intermediate GRBs seem to be distributed randomly, too.

\end{itemize}

\bigskip
\bigskip

We thank the valuable discussions with Drs. Michael Briggs,
Peter M\'e\-sz\'a\-ros, L\'aszl\'o P\'asztor, Dennis Sciama, G\'abor Tusn\'ady
and anonymous referee.
One of us (A.M.) thanks for the hospitality at Konkoly Observatory and
E\"{o}tv\"{o}s University. This article was partly supported
by GAUK grant 36/97, by GA\v{C}R grant 202/98/0522, by Domus
Hungarica Scientiarium et Artium grant (A.M.), by OTKA grant
T024027 (L.G.B) and by OTKA grant F029461 (I.H.).

\section*{FIGURES}

\includegraphics[height=150mm, width=0.90\textwidth]{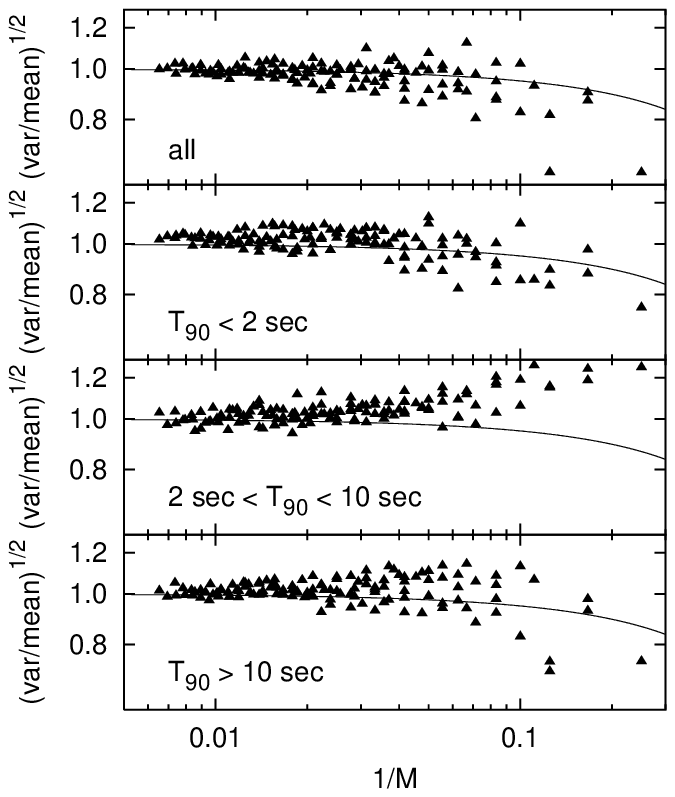}
\ \\
\ \\
Figure 1. The results of $105$ ``var/mean'' tests of four
different cases drawn in the $1/M$ vs. $\sqrt{\mbox{var/mean}}$ frame. The
theoretical curve
$\sqrt{1-1/M}$ (solid line) is also shown. $M$ is the number of cells.

\vfill \eject
\ \\
\includegraphics[height=90mm]{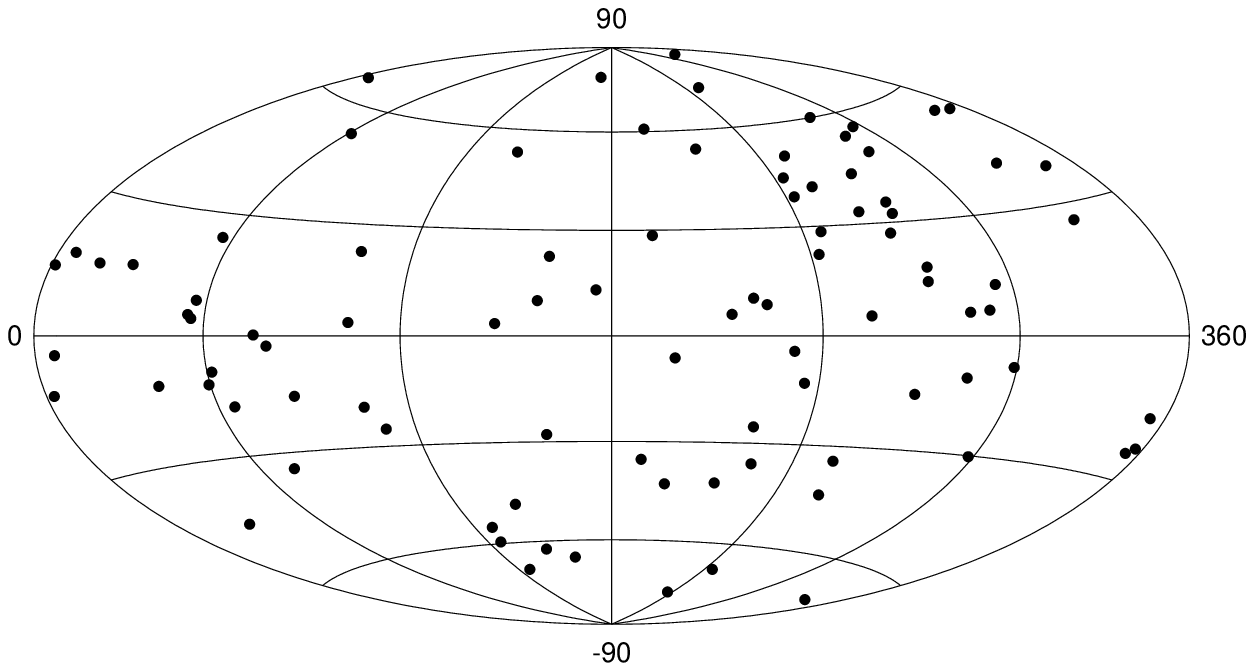} \ \\
Figure 2. Sky distribution of 92 GRBs of "dim" subclass of the 
intermediate subclass in equatorial coordinates.

\end{document}